\documentstyle[preprint,prd,aps]{revtex}
\tightenlines
\newcommand{\be}{\begin{equation}}
\newcommand{\ee}{\end{equation}}
\newcommand{\bea}{\begin{eqnarray}}
\newcommand{\eea}{\end{eqnarray}}
\begin{document}
\draft
\title{\large\bf  Twisted Boundary Conditions \\
and Matching to the \\
Effective Four Dimensional Theory}
\author{T.E. Clark\footnote{e-mail address: clark@physics.purdue.edu} and S.T. Love\footnote{e-mail address: love@physics.purdue.edu}}
\address{\it Department of Physics, 
Purdue University,
West Lafayette, IN 47907-1396}
\def\overlay#1#2{\setbox0=\hbox{#1}\setbox1=\hobx to \wd0{\hss #2\hss}#1%
\hskip -2\wd0\copy1}
\maketitle
\begin{abstract}
Nontrivial twisted boundary conditions associated with extra compact dimensions produce an ambiguity in the value of the four dimensional coupling constants of the renormalizable interactions of the twisted fields' zero modes. Resolving this indeterminancy  would require a knowledge of the exact form of the higher dimensional action including the coefficients of higher dimensional operators.  For the case of moderately sized extra dimensions, the uncertainty in the coupling constants can be of order one and may lead to modifications in the stability of the model.
\end{abstract}
\pacs{PACS number(s): 11.10.Kk, 11.25.Mj}
\vskip2pc
Theories with extra compact dimensions require that boundary conditions be specified on the
fields that depend on the bulk coordinates\cite{AMQABQ}-\cite{HoW}.  When viewed as an effective four dimensional field theory, the boundary conditions are reflected via an explicit phase dependence on the compact coordinates in the Fourier expansion coefficients of the fields. When the periodic boundary conditions involve field equality up to a global symmetry transformation phase, the additional coordinate derivative terms associated with the compact dimensions in the higher dimension Lagrangian will be replaced by a corresponding Kaluza-Klein mass factor in the four dimensional Lagrangian, even for the zero mode fields.  Indeed, by imposing periodic boundary conditions up to a $R$-transformation, Scherk and Schwarz provided a mechanism to break supersymmetry in four dimensions through dimensional compactification \cite{SS}-\cite{newSS}.   The purpose of this note is to point out that such nontrivial boundary conditions also lead to an ambiguity in the value of the coupling constants for interactions involving the zero modes of the twisted fields.  The ambiguity reflects the lack of knowledge of the exact value of the coefficients of the higher dimensional operators in the higher dimensional Lagrangian\cite{Hill}\cite{HK}.  When the compact dimensions are moderately sized compared to the inverse UV cutoff scale of the higher dimensional theory, the indeterminacy in the coupling constant can be of order one \cite{CL1} \cite{CL2}.

To specifically display this effect, we first consider the simplest example of a globally $U(1)$ invariant self interacting complex scalar field in five dimensions. Denoting the usual four dimensional space-time coordinates of  Minkowski space by $x^\mu$, while $0\leq y\leq 2\pi R$ is the coordinate of the compact fifth dimension, the action for the self-coupled complex scalar field, $\phi (x,y)$ is given by
\bea
I &=& \int d^4x \int_0^{2\pi R} dy \left\{ \partial_M \phi^\dagger\partial^M\phi 
-V^{(5)}(\phi^\dagger \phi) + \sum_{m,n, M_1,...,M_m, N_1,...,N_n} \frac{1}{\Lambda^{M+N+\frac{3}{2}(m+n)-5}}\right. \cr
& & \left. \chi^{(5)}_{M_1,...,M_m, N_1,...,N_n}(\frac{\phi^\dagger}{\Lambda^{3/2}}, \frac{\phi}{\Lambda^{3/2}})\partial^{(M_1)} \phi^\dagger \cdots \partial^{(M_m)} \phi^\dagger \partial^{(N_1)}\phi \cdots \partial^{(N_n)}\phi  \right\}.
\eea
The notation in the sum of this formula is somewhat cryptic and thus requires some further explanation. The sum is over all possible interactions given by higher dimensional  operators involving at least two derivatives. A typical term contains equal numbers of $\phi^\dagger$ and $\phi$ fields (consistent with the global $U(1)$ symmetry) with a various number of five dimensional spacetime derivatives acting on the fields. Thus, for example, $\partial^{(N_i)}\phi$ is a term with $N_i$ five dimensional spacetime derivatives acting on $\phi$. The total number of such derivatives acting on all the $(\phi^\dagger)$ $\phi$ fields is $(M=M_1+...+M_m)$ $N=N_1+...+N_n$. The coefficients, $\chi^{(5)}_{M_1,...M_m,N_1,...N_n}$, are dimensionless field dependent couplings containing the appropriate factors of the five dimensional metric tensor needed to make each such term in the sum a Lorentz scalar. The dependence of these higher derivative terms on the UV cutoff scale, $\Lambda$, of the five dimensional theory is explicitly indicated. (Recall that in five spacetime dimensions, the scalar field has canonical mass dimension 3/2). $V^{(5)}$ is the scalar field potential describing its derivative free self-interactions. 
Retaining the first few higher dimensional terms with the lowest powers of $1/\Lambda$, the action takes the form
\bea
I &=& \int d^4x \int_0^{2\pi R} dy \left\{ \partial_M \phi^\dagger\partial^M\phi -\mu^2 \phi^\dagger \phi -\frac{\lambda^{(5)}}{\Lambda}(\phi^\dagger \phi)^2  -\frac{\chi^{(5)}}{\Lambda^2}\partial^2 \phi^\dagger \partial^2 \phi \right.\cr
 & &\left. -\frac{\chi^{(5)}_1}{\Lambda^3}\phi^\dagger \phi \partial_M \phi^\dagger\partial^M \phi - \frac{\chi^{(5)}_2}{\Lambda^3}\frac{1}{2}\left[\phi^\dagger \partial_M \phi \phi^\dagger \partial^M \phi +\partial_M \phi^\dagger \phi  \partial^M \phi^\dagger \phi \right]
+ \cdots \right\}.
\label{Action2}
\eea
For simplicity, the potential is chosen to include a quadratic mass term with mass parameter $\mu^2$ and a quartic interaction with the dimensionless five dimensional coupling constant $\lambda^{(5)}$. Higher dimensional derivative interactions through order $\frac{1}{\Lambda^3}$ have been included with the dimensionless five dimensional coupling constants $\chi^{(5)}$, $\chi^{(5)}_1$ and $\chi^{(5)}_2$.

Imposing periodic boundary conditions up to an $U(1)$ transformation on the scalar field so that 
\be
\phi (x,y+2\pi R) =e^{2\pi i a} \phi (x,y) ~,
\ee 
with $a$ an arbitrary real number, it may be expanded in a Fourier series as
\bea
\phi (x,y) &=& \frac{1}{\sqrt{2\pi R}}\sum_{n=-\infty}^{+\infty} \phi_n(x) 
e^{i (n+a)\frac{y}{R}}\cr
 &=& \frac{1}{\sqrt{2\pi R}}e^{i a \frac{y}{R}} \varphi (x) + ... ,
\label{FS1}
\eea
where in the second equality the four dimensional zero mode field, $\varphi (x)\equiv \varphi_0(x)$, has been explicitly singled out of the modal sum.  The scalar field effective action in four dimensions is obtained by substituting (\ref{FS1}) into the action (\ref{Action2}).  Focusing on the pure zero mode piece of the action, it reduces to 
\bea
I &=& \int d^4x  \left\{ Z\partial_\mu \varphi^\dagger \partial^\mu \varphi -\left[ \mu^2 +\left( 1+\frac{a^2\chi}{(\Lambda R)^2}\right)a^2 M^2 \right]  \varphi^\dagger \varphi - \left[ \lambda -\frac{ a^2\chi_1}{(\Lambda R)^2} +\frac{a^2\chi_2}{(\Lambda R)^2}\right] (\varphi^\dagger \varphi)^2 
\right.\cr
& & \left.  \qquad\qquad-\frac{1}{M^2}\frac{ \chi}{(\Lambda R)^2} \partial^2 \varphi^\dagger \partial^2 \varphi -\frac{1}{M^2}\frac{\chi_1}{(\Lambda R)^2}\varphi^\dagger \varphi
 \partial_\mu \varphi^\dagger \partial^\mu \varphi 
\right. \cr
 & &\left. \qquad\qquad\qquad\qquad- \frac{1}{M^2}\left(\frac{ \chi_2}{(\Lambda R)^2}\right)  \frac{1}{2}\left( 
\varphi^\dagger \partial_M \varphi \varphi^\dagger \partial^M \varphi +
\partial_M \varphi^\dagger \varphi \partial^M \varphi^\dagger \varphi \right)
+ \cdots \right\},
\label{Action4}
\eea
where $M^2=\frac{1}{R^2}$ is the Kaluza-Klein mass squared and $Z=1+\frac{2a^2\chi}{(\Lambda R)^2}$. 
In obtaining this action, the various five dimensional coupling constants have been rescaled to yield their respective four dimensional form according to
\bea
\label{RES}
\lambda &=& \frac{\lambda^{(5)}}{2\pi R\Lambda}~;~\chi = \chi^{(5)} \cr
\chi_1 &=& \frac{\chi^{(5)}_1}{2\pi R\Lambda}~;~\chi_2 = \frac{\chi^{(5)}_2}{2\pi R\Lambda}.
\eea
We see that in the effective four dimension theory of the zero modes, the effective mass term, $ \mu^2 +\left( 1+\frac{a^2\chi}{(\Lambda R)^2}\right)a^2 M^2 $, and quartic coupling, $\lambda -\frac{ a^2\chi_1}{(\Lambda R)^2} +\frac{a^2\chi_2}{(\Lambda R)^2}$, depend upon the precise values of the underlying five dimensional coupling constants. Thus to match the two theories requires a precise knowledge of higher dimensional couplings in the five dimensional model. Furthermore, for modestly sized compact dimensions, $\Lambda R \sim 1-10$, these \lq\lq threshold" corrections to the four dimensional coupling constants maybe comparable in magnitude to the original coupling constants\cite{Hill}-\cite{CL2}.

There can be some striking implications of this observation. For instance, since the contribution to the mass term proportional to the Kaluza-Klein mass 
$M^2$ is always positive (it arises from a piece of the five dimensional kinetic energy), then depending on its magnitude relative to the mass parameter $\mu^2$, its presence could alter the sign of the overall mass term coefficient and hence modify the symmetry phase. In a similar vein, the relative signs and sizes of the coupling constants $\lambda$, $\chi_1$ and $\chi_2$ could effect the stability of the model.  Indeed, in order to insure vacuum stability, they must satisfy the inequality
\be
\lambda -\frac{a^2 \chi_1}{(\Lambda R)^2} +\frac{a^2 \chi_2}{(\Lambda R)^2} \geq 0 
\ee
in $(\lambda, \chi_1, \chi_2)$ coupling constant space.

Similar ambiguities arise in matching onto four dimensional theories containing fermions. Twisted fermion boundary conditions as well as twisted scalar boundary conditions can turn higher dimensional operators in the five dimensional theory into additional contributions to the fermion mass and effective Yukawa couplings involving the zero modes of the four dimensional theory.  Consider the simple example containing a six dimensional (eight component) five dimensional fermion $\Psi$, which consists of two four dimensional (four component) Dirac fermions in interaction with a hermitean scalar field, $\phi$, and  described by the five dimensional action
\bea
I &=& \int d^4x \int_0^{2\pi R} dy \left\{\frac{1}{2} \partial_M \phi \partial^M\phi -\mu^2 \phi^2 
-\frac{\lambda^{(5)}}{\Lambda} \phi^4 \right.\cr
 & &\left. +\frac{i}{2}\bar\Psi \Gamma^M \partial_M \Psi-(m_f +\frac{g^{(5)}}{\sqrt{\Lambda}}\phi)\bar\Psi \Psi \right\}.
\label{ActionF}
\eea
Since the scalar field is hermitean, it must obey untwisted boundary conditions. On the other hand, twisted boundary conditions for the fermions 
\be
\Psi (x,y+2\pi R) = e^{2\pi i a}\Psi (x,y) ,
\ee
can be enforced via its Fourier decomposition 
\bea
\Psi (x,y) &=& \frac{1}{\sqrt{2\pi R}}e^{i a \frac{y}{R}}\sum_{n=-\infty}^{+\infty}
\psi_n (x) e^{i n \frac{y}{R}} \cr
 &=& \frac{1}{\sqrt{2\pi R}}e^{i a \frac{y}{R}}\psi(x) + ... .
\eea
As before, in the second equality the eight component zero mode fermion field, $\psi(x)\equiv \psi_0(x)$, is singled out of the sum. If we now allow for the inclusion of higher dimensional operators in the five dimensional action such as 
\be
I_{\rm higher~dim}= \int d^4x \int_0^{2\pi R} dy \left\{ \frac{\chi^{(5)}_1}{\Lambda} \bar\Psi \partial^2 \Psi +\frac{\chi^{(5)}_2}{\Lambda^{5/2}}\phi \bar\Psi\partial^2 \Psi \right\}, 
\ee
then their presence produces additional contributions to the four dimensional fermion masses and Yukawa coupling constants for the zero modes. Explicitly, the mass squared becomes
\be
M_f^2 = m_f^2  +\left[ 1+ \frac{(a \chi_1)^2}{(\Lambda R)^2}\right] M^2,
\ee
while the four dimensional effective Yukawa coupling  becomes
\be
g_{\rm eff} = g + \frac{a^2 \chi_2}{(\Lambda R)^2}.
\ee
Here once again, these \lq\lq threshold effects" of the higher dimensional operators 
may be significant.

We conclude this note with an observation regarding the non-Abelian global symmetry breaking  structure of the zero mode four dimensional effective action arising from the twisted boundary conditions. The imposition of periodic boundary conditions up to a global symmetry transformation specifies a preferred direction in the symmetry group space and hence breaks the global symmetry.  Noether's theorem relates the variation of the Lagrangian, to the divergence of the current as
\be
\delta (\omega) {\cal L} = \partial_M J^M (\omega).
\ee
Here $\omega^i$ is the infinitesimal global transformation parameter, with $i$ enumerating  the number of parameters of the symmetry group, This in turn produces a variation of the action which depends on the four dimensional integral of the difference of the fourth component of the current which arises from one complete circuit around the compact dimension so that 
\be
\delta (\omega)  I = \int d^4x \left[ J^{M=4}(\omega)|_{y = 2\pi R} - J^{M=4}(\omega)|_{y = 0} \right].
\label{Noether}
\ee
For the case of a scalar field transforming according to a representation $T^i$ of the symmetry group, the twisted boundary conditions are of the form \cite{PQ}
\be
\phi (x, y+2\pi R) = e^{2\pi i \vec{a} \cdot \vec{T}} \phi (x,y),
\ee
with $\vec{a}$ a particular direction in group space. Satisfying this boundary condition leads to a zero mode decomposition of the form
\be
\phi (x,y) = \frac{1}{\sqrt{2\pi R}}U(y) \varphi (x) +... 
\ee
where $\varphi$ is the zero mode and $U(y)=e^{i\vec{a}\cdot \vec{T}\frac{y}{R}}$ is a group element. For definiteness, consider the model described by Eq.(\ref{Action2})  extended to allow for the non-Abelian global symmetry. The resultant action is 
\bea
I &=& \int d^4x \int_0^{2\pi R} dy \left\{ (\partial_M \phi^\dagger\partial^M\phi) -\mu^2 (\phi^\dagger \phi) -\frac{\lambda^{(5)}}{\Lambda}(\phi^\dagger \phi)^2 -\frac{\chi^{(5)}}{\Lambda^2}\left(\partial^2 \phi^\dagger \partial^2 \phi \right) \right. \cr
 & & \left.  -\frac{\chi^{(5)}_1}{\Lambda^3}(\phi^\dagger \phi) (\partial_M \phi^\dagger\partial^M\phi )  -\frac{\chi^{(5)}_2}{\Lambda^3}\frac{1}{2}\left[(\phi^\dagger \partial_M \phi)(\phi^\dagger \partial^M \phi)+(\partial_M \phi^\dagger \phi)(\partial^M \phi^\dagger \phi)\right]  \right.  \cr
 & & \left. \qquad\qquad\qquad\qquad- \frac{\chi^{(5)}_3}{\Lambda^3} \frac{1}{2}\left[(\partial_M \phi^\dagger \phi) (\phi^\dagger \partial^M \phi)\right]
+ \cdots \right\},
\label{Action3}
\eea
where we employ a notation so that each of the terms within a parentheses is a group singlet. Note that this action contains an additional coupling relative to the $U(1)$ example which arises from the different ways the internal indices on the scalar fields can be contracted to form a group singlet. 

A straightforward application of Noether's theorem produces the five dimension symmetry current, 
\bea
J^M (x,y) &=& -i\left( 1-\frac{\chi^{(5)}_1}{\Lambda^3} (\phi^\dagger \phi)\right) (\phi^\dagger \vec{\omega}\cdot \vec{T} \stackrel{\leftrightarrow}{\partial^M}\phi) \cr
 & &-\frac{i}{2} \left(2\frac{\chi^{(5)}_2}{\Lambda^3}-\frac{\chi^{(5)}_3}{\Lambda^3}\right) (\phi^\dagger
\vec{\omega}\cdot \vec{T} \phi)(\phi^\dagger \stackrel{\leftrightarrow}{\partial^M}\phi) \cr
 & &-i\frac{\chi^{(5)}}{\Lambda^2}\left[ (\partial^2 \phi^\dagger \vec{\omega}\cdot \vec{T} 
\stackrel{\leftrightarrow}{\partial^M}\phi) +(\phi^\dagger  \vec{\omega}\cdot \vec{T} 
\stackrel{\leftrightarrow}{\partial^M}\partial^2 \phi)\right].
\eea
The corresponding four dimensional zero mode contribution for its fourth component is given by
\bea
\label{J4}
\cr
J^{M=4}(\omega) &=& \frac{-iM}{2\pi }\left(1-\frac{1}{M^2}\frac{\chi_1}{(\Lambda R)^2}(\varphi^\dagger \varphi) \right)
(\varphi^\dagger  U^\dagger (y) \vec{\omega}\cdot \vec{T} \stackrel{\leftrightarrow}{\partial_y} U(y) \varphi )  \cr
 & &\frac{-i}{2\pi M}\frac{(2\chi_2-\chi_3)}{2(\Lambda R)^2}(\varphi^\dagger U^\dagger (y) \vec{\omega}\cdot \vec{T} U(y) \varphi)(\varphi^\dagger U^\dagger (y) 
\stackrel{\leftrightarrow}{\partial_y} U(y) \varphi ) \cr
 & & -\frac{i}{2\pi M}\frac{\chi}{ (\Lambda R)^2}\left[(  \partial^2\varphi^\dagger  U^\dagger (y) \vec{\omega}\cdot \vec{T}  \stackrel{\leftrightarrow}{\partial_y} U(y) \varphi) + (\varphi^\dagger
U^\dagger (y) \vec{\omega}\cdot \vec{T}  \stackrel{\leftrightarrow}{\partial_y} U(y) \partial^2 \varphi)\right] \cr
 &=&\frac{M^2}{2\pi }\left(1-\frac{1}{M^2}\frac{\chi_1}{(\Lambda R)^2}(\varphi^\dagger \varphi) \right) (\varphi^\dagger  ( 2\vec{\omega}\cdot\vec{a} +d^{ijk}\omega^i a^j U^\dagger (y) T^k U(y) ) \varphi )  \cr
& &+\frac{(2\chi_2-\chi_3)}{2\pi (\Lambda R)^2}(\varphi^\dagger U^\dagger (y) \vec{\omega}\cdot \vec{T} U(y) \varphi)(\varphi^\dagger \vec{a}\cdot \vec{T} \varphi ) \cr
& & +\frac{\chi}{2\pi (\Lambda R)^2}\left[( \varphi^\dagger [\stackrel{\leftarrow}{\partial^2} +M^2(\vec{a}\cdot\vec{T})^2]
 ( 2\vec{\omega}\cdot\vec{a} +d^{ijk}\omega^i a^j U^\dagger (y) T^k U(y) ) \varphi) \right.\cr
 & &\left. \qquad\qquad\qquad+ (\varphi^\dagger
( 2\vec{\omega}\cdot\vec{a} +d^{ijk}\omega^i a^j U^\dagger (y) T^k U(y) )
[\partial^2 + M^2(\vec{a}\cdot\vec{T})^2]\varphi)\right]  ,
\eea
where the $d^{ijk}$ arise through the anticommutator relation $\{T^i, T^j\}= 2\delta^{ij} + d^{ijk} T^k$. As was the case in the $U(1)$ example, we have introduced the effective four dimensional coupling constants via the same rescalings of their five dimensional versions as 
in Eq. (\ref{RES}) along with the rescaling $\chi_3 = \frac{\chi^{(5)}_3}{2\pi R\Lambda}$ and have used that the Kaluza-Klein mass is $M = 1/ R$.

Substituting Eq. (\ref{J4}) into Eq. (\ref{Noether}) yields the symmetry breaking boundary condition effects on the zero mode four dimensional action 
\bea
\delta (\omega) I &=& \int d^4 x  \frac{\omega^i}{2\pi} (e^{2\pi i \vec{a}\cdot \vec{t}}-1)_{ij} \cr
& &\left\{ \left(\vec{a}\cdot\vec{d}\right)_{jk}\left( M^2\left[\left(\varphi^\dagger T^k \varphi\right)  +\frac{\chi}{(\Lambda R)^2}\left(\varphi^\dagger \left\{T^k, \vec{a}\cdot\vec{T}\right\}\varphi \right) \right]\right.\right.\cr
& & \left.\left. -\left[\frac{\chi_1}{(\Lambda R)^2}\left(\varphi^\dagger \varphi\right)\left(\varphi^\dagger T^k \varphi\right) +\frac{2\chi}{(\Lambda R)^2} \left(\partial_\mu \varphi^\dagger T^k \partial^\mu \varphi \right) \right]\right) \right. \cr
& & \left. +\frac{(2\chi_2 -\chi_3)}{(\Lambda R)^2} \left(\varphi^\dagger T^j \varphi\right) \left(\varphi^\dagger \vec{a}\cdot \vec{T} \varphi \right)\right\} ,
\label{varAction44}
\eea
where $(t^j)_{ik} = i f_{ijk}$ are the adjoint representation matrices and $(\vec{a}\cdot\vec{d})_{ij}=a^k d^{ikj}$. In obtaining this result, the commutator relation $[ \vec{a}\cdot \vec{d }, \vec{a}\cdot \vec{t}] = 0$ has been exploited.  

As could have been anticipated there are breaking terms proportional to the Kaluza-Klein mass squared, $M^2$. On the other hand, there are also breaking terms which are dimension four operators which do not have an explicit $M^2$ coefficient. These arise from higher dimensional operators in the five dimensional action in which derivatives act on the compact dimension and once again reflect the type of effect we discussed earlier in the paper.

An alternative, equivalent, method of securing Eq. (\ref{varAction44}), is to first 
directly compute the zero mode four dimensional action by a  reduction of the five dimensional action, Eq. (\ref{Action3}). So doing, one finds
\bea
I &=& \int d^4 x \left\{ (\partial_\mu \varphi^\dagger [ 1+ \frac{2\chi}{(\Lambda R)^2}
(\vec{a}\cdot\vec{T})^2 ] \partial^\mu \varphi ) -(\varphi^\dagger [ \mu^2 + (\vec{a}\cdot\vec{T})^2 M^2 + \frac{\chi}{(\Lambda R)^2}(\vec{a}\cdot\vec{T})^4 M^2 ] \varphi )\right. \cr
 & &\left. -\lambda (\varphi^\dagger \varphi )^2 +\frac{\chi_1}{(\Lambda R)^2} (\varphi^\dagger \varphi ) (\varphi^\dagger (\vec{a}\cdot\vec{T})^2 \varphi ) -\frac{(2\chi_2 -\chi_3)}{(\Lambda R)^2}\frac{1}{2}(\varphi^\dagger \vec{a}\cdot\vec{T} \varphi )^2 \right. \cr
 & &\left. \qquad\qquad\qquad\qquad -\frac{1}{M^2}\frac{\chi}{(\Lambda R)^2}(\partial^2 \varphi^\dagger \partial^2 \varphi ) -\frac{1}{M^2}\frac{\chi_1}{(\Lambda R)^2}(\varphi^\dagger \varphi) (\partial_\mu \varphi^\dagger \partial^\mu \varphi ) \right. \cr
 & &\left.-\frac{1}{M^2}\frac{\chi_2}{2(\Lambda R)^2} \left[(\varphi^\dagger \partial_\mu \varphi)(\varphi^\dagger \partial^\mu \varphi)+(\partial_\mu \varphi^\dagger \varphi)(\partial^\mu \varphi^\dagger \varphi)\right]\right.\cr
 & & \left.-\frac{1}{M^2} \frac{\chi_3}{2(\Lambda R)^2} \left[(\partial_\mu \varphi^\dagger \varphi) (\varphi^\dagger \partial^\mu \varphi)\right] \right\}.
\label{Action44}
\eea
Once again, as a consequence of the twisted boundary conditions, certain higher dimensional terms in the five dimensional action which involve derivatives with respect to the compactified dimension become dimension four operators in this action. The explicit breaking terms are those which contain the specific group direction $\vec{a}$ picked out by the boundary conditions. 

The variation of this action can then be taken directly. One must be careful, however, to note that the twisted boundary conditions built in to the modal expansion lead to a transformation law for the zero mode fields which is boundary condition dependent and somewhat non-canonical.  Specifically, if the five dimensional fields transform as
\be
\delta (\omega) \phi (x,y) = i\vec{\omega}\cdot\vec{T} \phi (x,y) ,
\ee
then the zero mode varies as 
\be
\delta (\omega) \varphi (x) = i\vec{\omega}(a)\cdot\vec{T} \varphi (x) = i \omega^i \left[ \frac{e^{2\pi i\vec{a}\cdot\vec{t}} -1}{2\pi i \vec{a}\cdot\vec{t}}\right]_{ij} T^j \varphi (x) ,
\ee
which has the form of the usual symmetry transformation but with a boundary condition dependent parameter. Using this variation explicitly in Eq. (\ref{Action44}), one regains the transformation of the action as given in Eq. (\ref{varAction44}).

\noindent
This work was supported in part by the U.S. Department of Energy under grant DE-FG02-91ER40681 (Task B).

\end{document}